\documentclass[fleqn,usenatbib]{mnras}
\usepackage{hyperref}
\usepackage{newtxtext,newtxmath,times,verbatim}
\usepackage[T1]{fontenc}
\usepackage{ae,aecompl}
\usepackage{graphicx}
\usepackage{amsmath}
\usepackage{amssymb}
\usepackage{commath}
\usepackage{comment}
\DeclareGraphicsExtensions{.png,.pdf}
\usepackage{xfrac}
\DeclareMathOperator{\atantwo}{arctan2}

\author[Uzeirbegovic et al.]{Emir
  Uzeirbegovic,\thanks{e.uzeirbegovic@herts.ac.uk} James~E.~Geach and
  Sugata Kaviraj\\
Centre for Astrophysics Research, School of Physics, Astronomy \&
Mathematics, University of Hertfordshire, Hatfield, AL10 9AB\\
Centre of Data Innovation Research, School of Physics, Astronomy \&
Mathematics, University of Hertfordshire, Hatfield, AL10 9AB}

\pubyear{2020}

\title[Eigengalaxies]{Eigengalaxies: describing galaxy morphology
  using principal components in image space}

\begin{document}
\label{firstpage}
\pagerange{\pageref{firstpage}--\pageref{lastpage}}
\maketitle
\begin{abstract}
We demonstrate how galaxy morphologies can be represented by weighted sums of `eigengalaxies' and how eigengalaxies can be used in a probabilistic framework to enable principled and simplified approaches in a variety of applications. Eigengalaxies can be derived from a Principal Component Analysis (PCA) of sets of single- or multi-band images. They encode the image space equivalent of basis vectors that can be combined to describe the structural properties of large samples of galaxies in a massively reduced manner. As an illustration, we show how a sample of 10,243 galaxies in the {\it Hubble Space Telescope} CANDELS survey can be represented by just 12 eigengalaxies. We show in some detail how this image space may be derived and tested. We also describe a probabilistic extension to PCA (PPCA) which enables the eigengalaxy framework to assign probabilities to galaxies. We present four practical applications of the probabilistic eigengalaxy framework that are particularly relevant for the next generation of large imaging surveys: we (i) show how low likelihood galaxies make for natural candidates for outlier detection (ii) demonstrate how missing data can be predicted (iii) show how a similarity search can be performed on exemplars (iv) demonstrate how unsupervised clustering of objects can be implemented. 

\end{abstract}

\section{Introduction}

The distribution of light in galaxies, commonly referred to as galaxy
`morphology', is a fundamental observable property. Morphology
strongly correlates with the physical properties of a galaxy, such as
its stellar mass \citep[e.g.][]{Bundy2005}, star formation rate
\citep[e.g.][]{Ryan2012,Bluck2014,Smethurst2015,Willett2015}, surface
brightness \citep[e.g.][]{Martin2019,Jackson2020b}, rest frame colour
\citep[e.g.][]{Strateva2001,Skibba2009,Bamford2009} and local
environment \citep[e.g.][]{Dressler1997,Postman2005} and reveals key
information about the processes that have shaped its evolution over
cosmic time \citep[e.g.][]{Martin2018b,Jackson2020}. For example, the
smooth light distributions of elliptical galaxies, which are a result
of the largely random orbits of their stars
\citep[e.g.][]{Cappellari2011}, are signposts of a merger-rich
evolutionary history \citep[e.g.][]{Conselice2006}. On the other hand,
the presence of a disc indicates a relatively quiescent formation
history, in which the galaxy has grown primarily through accretion of
gas from the cosmic web \citep{Codis2012,Martin2018c}. In a similar
vein, morphological details such as extended tidal features suggest
recent mergers and/or interactions
\citep[e.g.][]{Kaviraj2014b,Kaviraj2019,Jackson2019}, with the surface
brightness of these tidal features typically scaling with the mass
ratios of the mergers in question
\citep[e.g.][]{Peirani2010,Kaviraj2010}. 

Apart from being a fundamental component of galaxy evolution studies,
morphological information has a wide range of applications across
astrophysical science. For example, it can be a key prior in
photometric redshift pipelines \citep[e.g.][]{Soo2018,Menou2018} which
underpin much of observational cosmology and weak lensing studies, is
used as contextual data in the classification of transient light
curves \citep[e.g.][]{Djorgovski2012,Wollaeger2018} and is an
essential ingredient in the study of the processes that drive active
galactic nuclei \citep[e.g.][]{Schawinski2014,Kaviraj2015}. The
morphological analysis of galaxy populations, especially in the large
surveys that underpin our statistical understanding of galaxy
evolution, is therefore of fundamental importance.

A vast literature exists on methods for measuring galaxy
morphology. Popular techniques range from those that describe a
galaxy's light distribution using a small number of parameters
\citep[e.g.][]{deVaucouleurs1948,Sersic1963,
  Simard2002,Odewahn2002,Lackner2012}
to non-parametric approaches such as `CAS'
\citep[e.g.][]{Abraham1994,Conselice2003,Menanteau2006,Mager2018} or
Gini-$M_{20}$ \citep[e.g.][]{Lotz2004,Scarlata2007,Peth2016}, where
the light distribution is reduced to a single value. The convergence
of large observational surveys and rapidly increasing computing power
has recently brought machine learning to the fore in morphological
studies. While the use of machine learning can be traced back at least
as far as the 1990s \citep[e.g.][]{Lahav1995}, there has been a
recent explosion of studies that apply such techniques to the
exploration of galaxy morphology, particularly in large survey
datasets
\citep[e.g.][]{Huertas-Company2015,Ostrovski2017,Schawinski2017,
  Hocking2018, Goulding2018,Cheng2019,Martin2020}.

Automated techniques lend themselves particularly well to the analysis
of large surveys, but the most accurate method of morphological
classification is arguably visual inspection
\citep[e.g.][]{Kaviraj2007,Kaviraj2019}. Indeed the genesis of this subject can be
traced back to the visual `tuning-fork' classifications by
\citet{Hubble1926}, where galaxies were classified into a so-called
sequence of ellipticals and spirals. It is remarkable that this
classification system still underpins the broad morphological classes
into which galaxies are split in modern studies of galaxy
evolution. Although visual inspection of large observational surveys
is time-intensive, the advent of massively distributed systems like
Galaxy Zoo has revolutionised its use on survey datasets
\citep[e.g.][]{Lintott2011,Simmons2014,Simmons2017,Willett2017}. Galaxy Zoo has
used more than a million citizen-science volunteers to classify large
contemporary surveys, like the Sloan Digital Sky Survey (SDSS) and the
{\it Hubble Space Telescope} ({\it HST}) Legacy Surveys and has
provided the benchmark against which the accuracy of automated
techniques have been routinely compared
\citep[e.g.][]{Huertas-Company2015,Dieleman2015,Beck2018,Walmsley2019,Ma2019}.

In contrast to classification based approaches, this paper describes how 
morphology may be interpreted as the `layout' of an image space which is survey specific, 
linear and has a probabilistic interpretation. We show how these properties enable the 
same image space to serve as a multi-application framework.
The paper is structured as follows. In Section \ref{intro} we describe
both the sample and our methodology, we discuss the construction of the image space,
its validity and its relevance to the interpretation of galaxy morphology. We also show
how this result is extended to include a probabilistic interpretation. In Section
\ref{applications} we show how this framework leads to principled approaches in various
applications. In particular, we (i) show how low likelihood galaxies make for natural
candidates for outlier detection (ii) show how missing data can be predicted (iii) 
demonstrate how similarity searches for images can be implemented given an exemplar
(iv) demonstrate how a natural unsupervised clustering of objects can be produced. 
We then outline how the methods may be used together in more advanced applications and
how it is relevant to big datasets such as the upcoming Large Synoptic Survey Telescope
(LSST) project \citep{Robertson2017}. Section \ref{conclusion} concludes and summarises 
our findings, and provides a link to the codes developed for this work.

\section{Eigengalaxies}\label{intro}

\begin{figure*}
\centering
\includegraphics[width=0.16\textwidth]{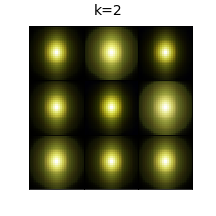}
\includegraphics[width=0.16\textwidth]{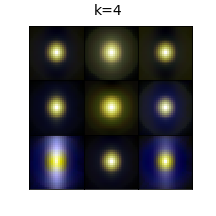}
\includegraphics[width=0.16\textwidth]{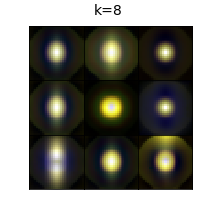}
\includegraphics[width=0.16\textwidth]{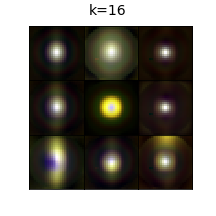}
\includegraphics[width=0.16\textwidth]{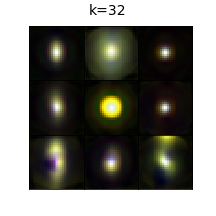}
\includegraphics[width=0.16\textwidth]{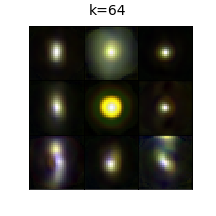}
\includegraphics[width=0.16\textwidth]{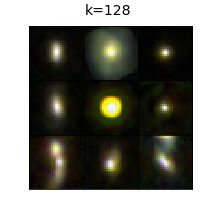}
\includegraphics[width=0.16\textwidth]{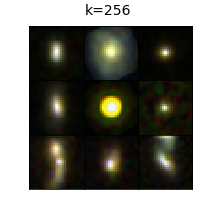}
\includegraphics[width=0.16\textwidth]{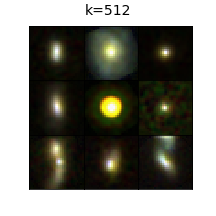}
\includegraphics[width=0.16\textwidth]{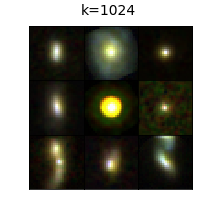}
\includegraphics[width=0.16\textwidth]{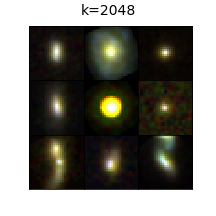}
\includegraphics[width=0.16\textwidth]{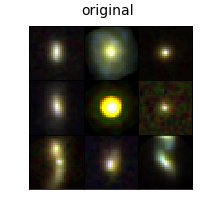}

\caption{
  RGB composite (F160W, F125W, and F814W) thumbnails
  for a selection of galaxies reconstructed using different numbers of
  eigengalaxies (indicated by $k$). The number of eigengalaxies
  increases by powers of $2$. The original image is shown in the
  bottom right. The fidelity of the reconstruction is directly related
  to the retention of eigengalaxies. Experimentation reveals that
  $12$ eigengalaxies were enough to robustly preserve neighbourhood
  similarity.
} 
\label{fig:recon}
\end{figure*}

\subsection{Data from {\it HST} CANDELS}\label{data}

{\it HST} CANDELS \citep{grogin2011candels,koekemoer2011candels}
offers a high-resolution probe of galaxy evolution. The survey
consists of optical and near-infrared (WFC3/UVIS/IR) images from the
Wide Field Camera 3 (WFC3) and optical images from the Advanced Camera
for Surveys (ACS) in five well-studied extragalactic survey
fields. Here, we focus on GOODS-S, one of the deep tier (at least
four-orbit effective depth) fields. We select a sample of 10,243
galaxies present in the `Galaxy Zoo: CANDELS' (GZ-CANDELS) GOODS-S
catalogue \citep{simmons2016galaxy}, that fall within the region
jointly covered by the F814W, F125W and F160W bands. The majority of
objects are at $z<3$ \citep{simmons2016galaxy}.

For each object in the catalog we take a 1.8$''$ cut-out, using the
catalogued sky coordinate as a centroid. ACS images are down-sampled
by a factor of two to match the pixel scale of the WFC3 images
(0.06$''$ pixel$^{-1}$), so that for each band we obtain a $30\times
30$-pixel image at the position of each galaxy. The cut-out dimensions
are selected with prior experimentation because they result in the
least superfluous background for most targets. Individual cut-outs
are given as $30\times 30\times 3$ pixel array in which the dimensions
index the height, width and band respectively.

\subsection{Morphology as an image space}\label{imagespace}

We may gain considerable flexibility by thinking about galaxies as
points in a vector space and about morphology as the `layout' of the
space. In this section we describe a simple vector space sufficient
for our demonstrations in Section \ref{applications}.

The first task is to construct a projection from the $30\times
30\times 3$ cut-outs to vectors in a $j$-dimensional space. 
This kind of embedding has been done in many ways in 
existing literature. Some examples from astronomical 
applications include \cite{naim1997galaxy} who pre-calculate 
features thought to be important in order to feed a self-organising map
\citep{kohonen1990self}, and \cite{Hocking2018} who uses patches of
pixels as a primitive and then calculate histograms of patches to
reduce images to vectors. Here, we avoid techniques which
presuppose meaningful summarisations from the outset and begin by
making a naive projection by flattening cut-outs into
vectors\footnote{The order in which the array is flattened does not
  matter because we will only be interested in the pairwise 
  relationships between pixels, which is not changed if they are
  presented in a different order.}, hence $j=2700$. There are many
possible constraints and coherences that may be desirable in an image
space but we will be singularly interested in preserving the
similarity of galaxy points which are `near' each other. This
generates clear testable implications which we can validate. The two
main factors bearing on this correspondence are how the projection
maps galaxies to vectors and how the distance between matrices is
calculated. These will now be considered in turn.

A first necessary modification to the naive projection is to remove
uninteresting sources of variation. In addition to the image
centering, pixel scaling and cropping described in Section \ref{data},
we identify range clipping and rotation to a common plane as the most
important modifications. To achieve these we truncate the pixel
distribution at the 99$^\text{th}$ percentile to avoid extreme
spikes in flux in some galaxies from making them seem excessively
different to the rest. Next, we produce a temporary composite image of
the array by using, for each pixel, its maximum magnitude across all
bands. We then use the locations of all pixels in our composite with
values exceeding the 75$^\text{th}$ percentile to create a matrix of
the coordinates of bright pixels. The first principal component of the
two column matrix of pixel coordinates is a vector of weights
$(x_0,y_0)$ which can be interpreted as the linear transformation of
our 2D coordinates to the 1D space that preserves most variance
(i.e.\ some line passing through the origin of the original 2D
space). The angle between the original {\it x}-axis in our image and
the new variance-preserving one is then given by $\theta
=\atantwo(y_0,x_0)$. Finally, we transform the original array by
rotating every band in turn by $\theta -\pi/2$ radians (in order to
make vertical what would otherwise be a rotation to the horizontal
plane) which results in vertically-aligned brightness.

\subsection{Low-rank approximation} \label{lowrank}

\begin{figure*}
\centering
\includegraphics[width=0.16\textwidth]{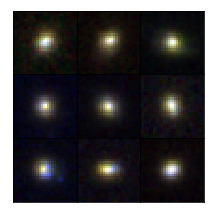}
\includegraphics[width=0.16\textwidth]{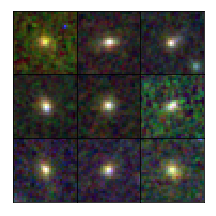}
\includegraphics[width=0.16\textwidth]{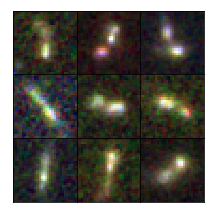}
\includegraphics[width=0.16\textwidth]{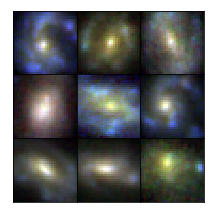}
\includegraphics[width=0.16\textwidth]{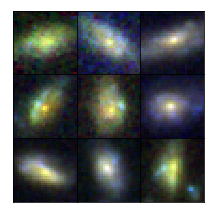}
\includegraphics[width=0.16\textwidth]{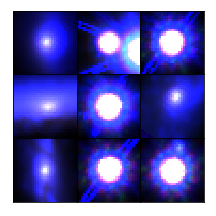}
\includegraphics[width=0.16\textwidth]{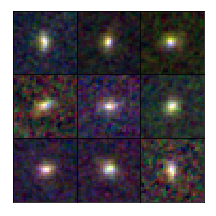}
\includegraphics[width=0.16\textwidth]{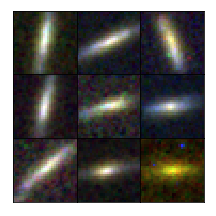}
\includegraphics[width=0.16\textwidth]{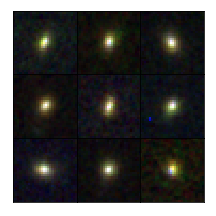}
\includegraphics[width=0.16\textwidth]{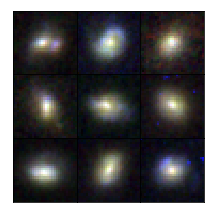}
\includegraphics[width=0.16\textwidth]{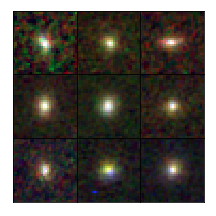}
\includegraphics[width=0.16\textwidth]{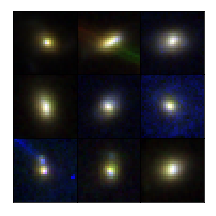}
\includegraphics[width=0.16\textwidth]{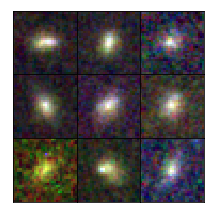}
\includegraphics[width=0.16\textwidth]{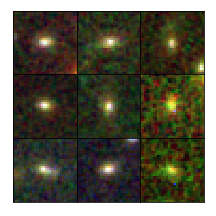}
\includegraphics[width=0.16\textwidth]{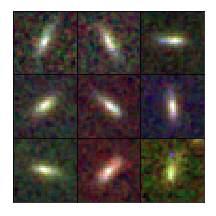}
\includegraphics[width=0.16\textwidth]{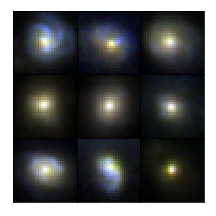}
\includegraphics[width=0.16\textwidth]{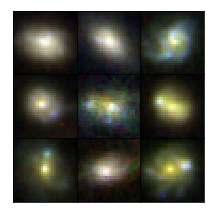}
\includegraphics[width=0.16\textwidth]{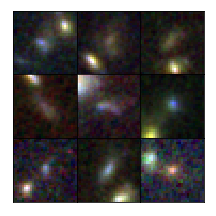}

\caption{
  A collection of RGB composite (F160W, F125W, and
  F814W) thumbnail sets, in each of which the galaxy in the top left is
  randomly selected and is followed by 8 of the nearest galaxies to it
  in image space, ordered by row. Even for noisy and complicated
  examples, the nearest neighbours tend to be visually similar
  suggesting that the image space does a good job of preserving visual
  similarity in local neighbourhoods.
} 
\label{fig:similar-sample}
\end{figure*}

Having fixed how images are to be represented in our projection, we
can further improve the image space by noticing that flux density in
image pixels are typically highly correlated. This implies that it may
be possible to closely approximate a galaxy vector in fewer than $j$
dimensions. This is very much desirable because a $j=2700$ dimensional
space will likely suffer from some of the degeneracies sometimes
termed `the curse of dimensionality' which includes effects like the
loss of interpretability of distance measures discussed further in Section \ref{bigdata}.

Let $S=(v_1,...,v_L)$ be a survey matrix in which the row vectors
$v_i$ with $j$ components are the projected images. We are interested
in $\hat{S}$ with dimensions $L\times k$ where $k<j$. The associated
minimisation problem can be expressed as:

\begin{equation} \label{mineq}
  \min_{\hat{S}} ||S-\hat{S}|| \text{ s.t. }rank(\hat{S})< l.
\end{equation}

Here, $||\cdot||$ (the norm) governs how the distance between vectors
is quantified. The Frobenius norm \citep{gloub1996matrix} is given by:

\begin{equation}
  ||X||=\sqrt{\sum_{i=0}^m\sum_{j=0}^n |x_{i,j}|^2}.
\end{equation}

This is commonly used and can be thought of as a matrix generalisation
of the Euclidean distance. Given a Frobenius norm, the minimisation
problem in equation \ref{mineq} has a globally optimal, unique and
analytical solution as a consequence of the Eckart-Young-Mirsky (EYM)
theorem \citep{eckart1936}.  The singular value decomposition of $S$
is such that $S=U\Sigma V^T$ where $U,V$ are orthogonal matrices and
$\Sigma$ is a diagonal matrix with $(\sigma_1,...,\sigma_k)$ singular
values. The theorem shows that given a Frobenius norm, the optimal and
unique solution to equation \ref{mineq} for a matrix $\hat{S}$ of rank
$k$ is given by  $\hat{S}=\sum_{i=1}^k \sigma_i u_i v_i^T$, where
$k\le l$, and $u_i,v_i$  are vectors from $U,V$ respectively.

Principal component analysis (PCA) is a recursive procedure in which
a plane (principal component, PC) which minimises the average squared 
distance from points to the PC is fitted to the residuals of the previous
iteration. The process carries on until there are no further
residuals. An efficient way to measure the contribution of each PC is
to examine the covariance matrix of the data projected onto the new
basis. Since each PC is required to be orthogonal, the covariance
matrix is diagonal. Further, since residual difference is smaller
on each iteration, subsequent PCs account for progressively less
variance thus creating a natural ordering. The cumulative sum of the
diagonal elements of the covariance matrix therefore provides a
convenient way to measure the total fraction of variance
accounted for by $k$ PCs. This measure is often termed ``explained
variance'' (EV) and the fraction of variance explained by any given PC
is termed the ``explained variance ratio'' (EVR). If the data is
centered, it turns out that PCs discovered in the iterative process
above are equivalent to the eigenvectors of the covariance matrix
$S^TS$ when ordered by their corresponding eigenvalues. It is further
the case that the eigenvectors are equivalent to the right singular
vectors in matrix $V$ above and the eigenvalues are given by
$\lambda_i=\sigma_i^2$. Thus, it can be shown by the EYM theorem that 
PCA is also an optimal solution to the minimisation problem posed in
equation \ref{mineq}.

PCA has a long precedence in astronomy applied in other ways to other
things. For example, \cite{de2004machine} use PCA to project a set of
310 disparate images to a lower rank space to facilitate further
classification steps. They referred to the basis eigenvectors as
`eigengalaxies' and used their weightings to test multiple machine
learning methods against each other for the purpose of galaxy
classification. \cite{li2005empirical} use PCA to decompose stellar
spectra from the STELIB spectroscopic stellar library
\citep{le2003stelib} and Two Micron All Sky Survey
\citep[2MASS,][]{skrutskie2006two} near-infrared photometry to derive
`eigenspectra' and then used them to fit the observed spectra of a
selection of galaxies from the SDSS \citep{fukugita1996sloan} Data
Release 1. \citet{anderson2004fast} generate many galaxies from a
parametric model, calculated the eigenvectors of the generated set,
and then used the reduced eigenspace to find the nearest synthetic
model. \citet{wild2014new} use PCA to concisely describe a large
number of model spectral energy distributions (SEDs). They termed the 
retained eigenvectors as `super-colours' and used them to impute SEDs
from sparse samples in observed galaxies. \cite{galaz1998eso} use PCA
to perform spectral classification of selected galaxies from the
ESO-Sculptor galaxy redshift survey data \citep{de1993mapping}.

In the analysis that follows we use the Frobenius norm and its vector 
equivalent, the Euclidean distance, as difference measures. We use PCA to 
optimally reduce the dimensionality of the image space. We term the 
eigenvectors as `eigengalaxies' because they are $j$ dimensional and can
be reshaped back into $30\times 30\times 3$ cut-outs which map how they 
are weighted as later shown in Figure \ref{fig:components}.

\subsection{Results and interpretations} \label{limitations}

The projection and low-rank approximation enables us to create a
low dimensional image space in which morphology is expressed in terms of
the distances between galaxies. The same general methodology may be used to
make image spaces for a broad range of surveys, and there are no strict
limitations on image sizes, bands or resolutions required. However, it is essential
to test that the projection and the low rank approximation result in a space fit for
purpose. Generally, PCA only fails when there are multiple identical eigenvalues
which results in the associated eigenvectors not being unique, rendering the result
meaningless. If PCA does not fail, it does not mean that it is appropriate to the 
problem at hand. However, there are a variety of heuristic means we can adopt to 
test the adequacy of the resultant space which we now cover.

Figure \ref{fig:recon} helps clarify what features of images the
space preserves. The figure shows composite thumbnails of a random
selection of galaxies followed by their reconstruction at different numbers
of eigengalaxies. Since our space aims to preserve
neighbourhood similarity, we can heuristically evaluate the
suitability of the space for any given $k$ by picking random galaxies
and then looking to see what their nearest neighbours look
like. Experimentation indicates that $k=12$ is enough for good
results. Figure \ref{fig:similar-sample} illustrates $18$ sets of
galaxies. In each set, the galaxy in the top left is a randomly
selected, followed by $8$ of the nearest galaxies to it in image
space, ordered by row. Even for noisy and complicated examples the
nearest neighbours tend to be visually similar, suggesting that the
image space does a good job of preserving visual similarity in local
neighbourhoods.

\begin{figure}
\centering
\includegraphics[scale=.55]{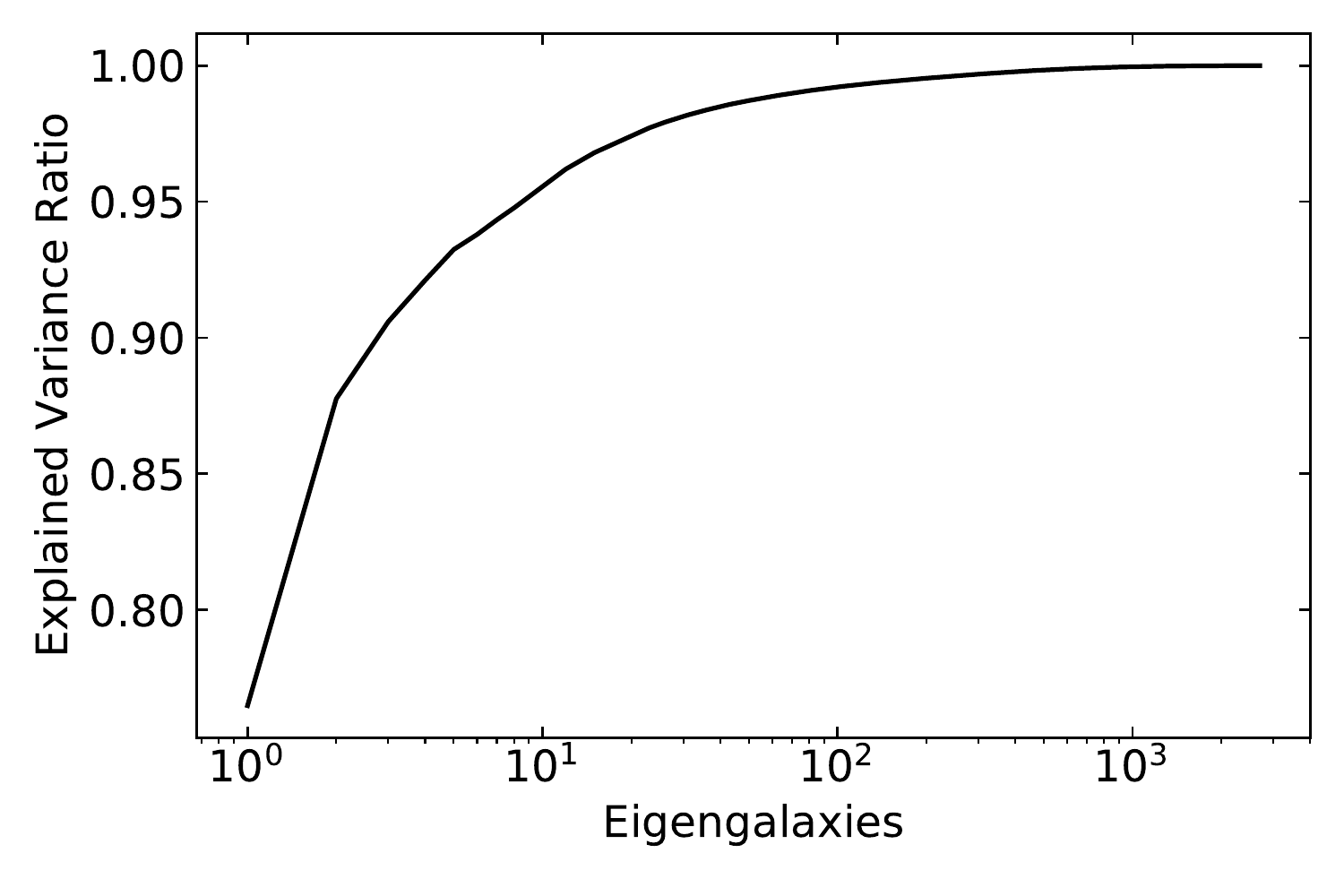}
\caption{Cumulative explained variance illustrating how much variance
  each additional feature accounts for. In PCA successive components
  account for less variance than the previous component. In this
  instance, only $2$ eigengalaxies are required to account for
  $\sim$85\% of explained variance yet $12$ eigengalaxies are required
  to account for $96\%$ of explained variance.}
\label{fig:cum-var}
\end{figure}

\begin{figure}
\centering
\includegraphics[scale=.73]{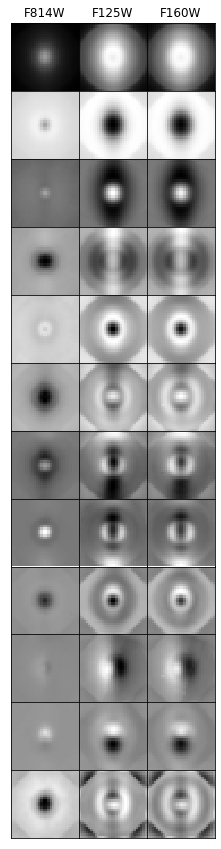}
\caption{The $12$ eigengalaxies accounting for $96\%$ variance in the
  GZ-CANDELS GOODS-S sample. Each row is an eigengalaxy and each
  column is a band. All images are scaled identically, where white
  indicates relative emphasis and black indicates relative de-emphasis
  of the region. Each image is $30\times30$ pixels, or $1.8''\times
  1.8''$.}
\label{fig:components}
\end{figure}

Figure~\ref{fig:cum-var} shows the cumulative variance explained by
successive eigengalaxies. PCA successively maximises the variance in
each orthogonal eigengalaxy and each additional eigengalaxy therefore
accounts for less variance. In this instance, only $2$
eigengalaxies are required to account for $\sim$85\% of explained
variance, but $12$ eigengalaxies are required for reliable nearest
neighbour similarity which accounts for $96\%$ of explained
variance. Since eigenvectors are $2700$ dimensional, we can
investigate this front loaded distribution by reshaping them back
into a $30\times 30\times 3$ images to get a sense of what each
eigenvector emphasises. Figure~\ref{fig:components} presents the $12$
eigengalaxies as grayscale images. It is evident that the first $2$
eigengalaxies focus on central brightness whilst later eigengalaxies
are successively more complicated, and therefore more likely
associated to the details of spatial correlation. The EVR directly
relates distances in the image space to the source of their variation,
hence morphology is herein described mostly by relative flux densities
across bands and coarse pixel correlation. 

\begin{figure}
\centering
\includegraphics[scale=.55]{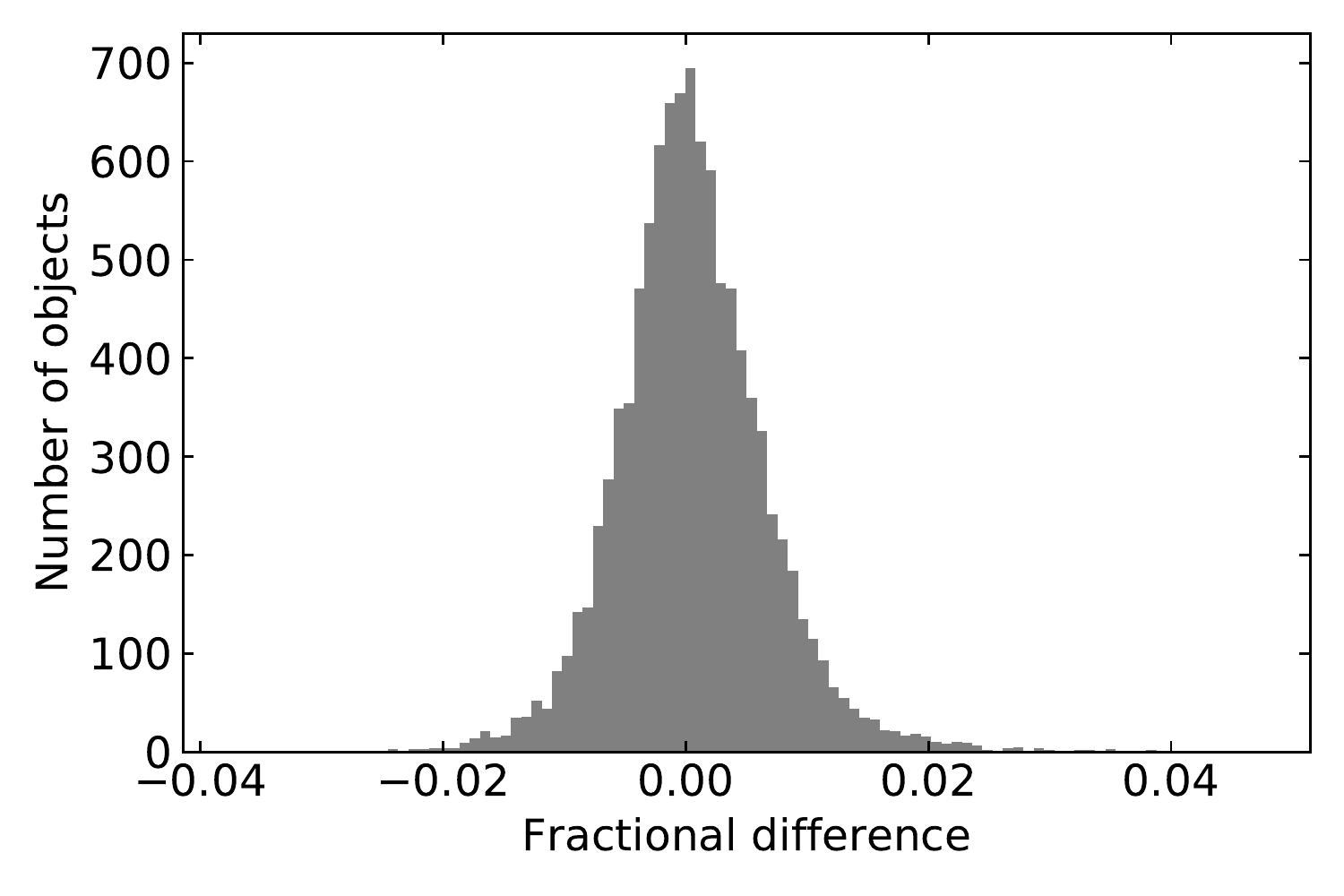}
\caption{Histogram showing the frequency of the ratio of the sum of
  deltas (difference between original and reconstructed pixel flux
  over all bands) and the sum of the flux in the original images, for
  the PCA decomposition of the 10,243 galaxies in the GZ-CANDELS
  GOODS--S sample. $99\%$ of objects lie between $\pm 2.5\%$. We account
  for $96\%$ of variance by design and the distribution above shows that
  the reconstructed images bear a good likeness (in terms of
  reconstructed flux) to the originals.}
\label{fig:flux-diff-hist}
\end{figure}

While we have assumed that the eigengalaxies are equally representative of
all galaxies, we can validate this assumption by graphically
checking the distribution of the reconstruction error, which is calculated
as the residual between the real image and its eigengalaxy based
reconstruction. If it is unevenly distributed or bigger than $4\%$
(i.e. $100\%-96\%$) then there may be something degenerate about the
image space. Let $F_0$ be the collection of flux densities of the
pixels in the original images and $F_\Delta$ be the corresponding
differences in flux between the original pixels and those in images
reconstructed using the eigengalaxies. Figure~\ref{fig:flux-diff-hist}
shows a histogram of \smash{$\sum_{x \in F_\Delta} x / \sum_{x \in
F_0} x$} (i.e.\ the ratio of the sum of deltas and the sum of flux
over all bands). It shows a symmetrical distribution centred at
zero. 99\% of objects lie between \smash{$\mid \sum_{x \in F_\Delta} x
  / \sum_{x \in F_0} x \mid$} $< 2.5\%$, hence it suggests that the
eigengalaxies do represent most galaxies similarly.

\begin{figure}
\centering
\includegraphics[scale=.55]{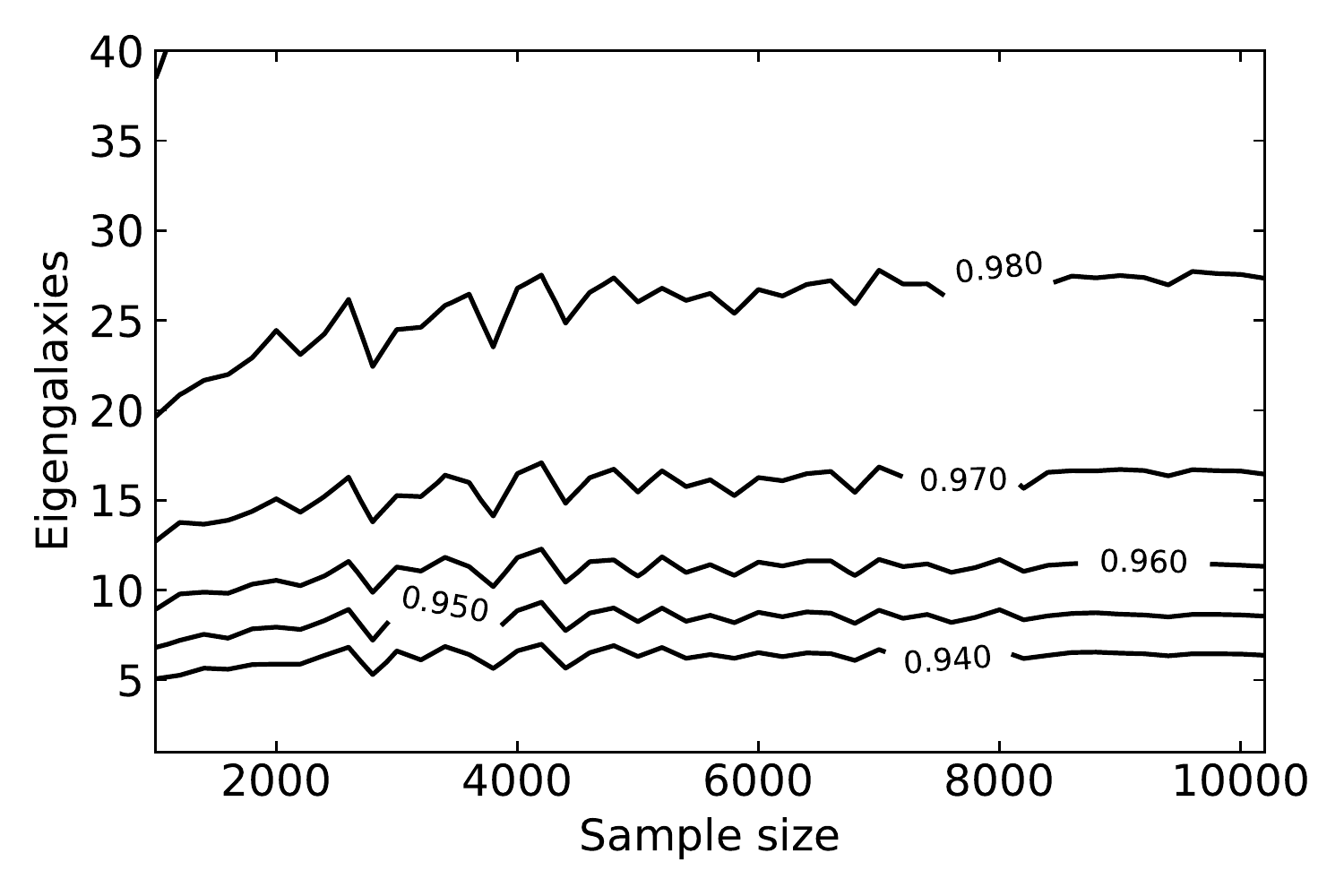}
\caption{Contour plot of explained variance corresponding to the
  required number of eigengalaxies versus sample size. Higher
  explained variances tend to require proportionally more
  eigengalaxies as sample size increases, however at $12$ eigengalaxies 
  explained variance is asymptotic to a constant.} 
\label{fig:pca-over-samples}
\end{figure}

We also assume that the eigengalaxies discovered are not an accident
of the sample size. We can check this by making sure that the same
number of eigengalaxies realise a similar explained variance at
smaller sample sizes and that therefore there is a reason to believe
that 12 eigengalaxies has an asymptotic
EVR. Figure~\ref{fig:pca-over-samples} presents a contour plot of EVR
corresponding to various combinations of sample size and eigengalaxy
number. We see that at 12 eigengalaxies the contour has near zero or
asymptotic growth as sample size increases, which suggests that the
vector space is less likely to be an accident of the sample. 

\begin{figure}
  \centering
  \includegraphics[width=8cm]{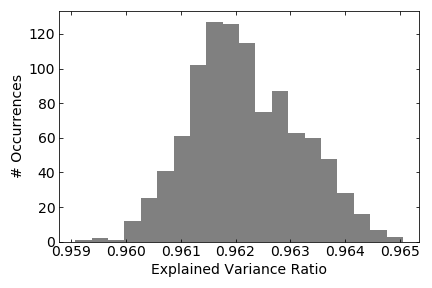}
  \includegraphics[width=8cm]{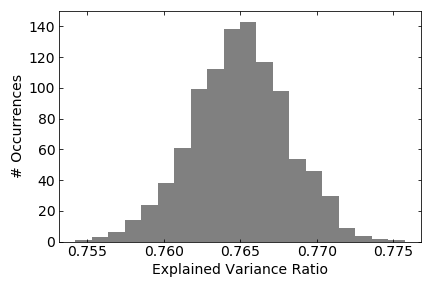}
  \caption {Both graphs show histograms of 1000 PCA fittings on 70\% of
    the data randomly sampled on each iteration. The top image shows the EVR for $k=12$. Note that the EVR at
    $k=12$ when the whole data is retained is $\sim 0.962$. The bottom
    image shows the EVR for the first eigenvector. Note that the EVR
    at $k=1$ when the whole data is retained is $\sim 0.765$. Both
    histograms show a very tight variance even at the extremes. }
  \label{bootstrap}
\end{figure}

PCA is sensitive to outliers, and can perform poorly where data has
very well separated clusters, and/or large non-linear sub-spaces. The
most likely cause of outliers in our case would be extreme brightness
against which we guard by range clipping. Further, there is no obvious
reason to expect large non-linear sub-spaces and we would expect
accidental well separated clusters to be highly unlikely in a 2700 dimensional
space. At the time of writing, there are no standard well known tests
which are able to diagnose these degenerate cases in high dimensional
spaces but it should be possible to identify their effects at least
heuristically by resampling the data, each time leaving out a large
part of it, and then calculating the EVR for some fixed $k$. 

We can pick $k=12$ which does a good job as described above and $k=1$, 
since the first eigenvector explains the most variance. If our data is
summarised robustly by PCA, the variance of the EVRs given $k$
components should be small. A large variance would give us reason
to believe that that the incidental presence/absence of outliers,
clusters, non-linear sub-spaces, or some other degeneracy is causing
substantial differences. It is important that the sample size taken on
every iteration is large enough to make catching these effects
likely. Figure \ref{bootstrap} illustrates a histogram of the EVR
values achieved by performing PCA on 1000 random 70\% samples of the
data for $k=12$ (top) and $k=1$ (bottom) respectively. The figure
shows distributions with tight ranges and thus provides some
confidence that the data is appropriate for PCA as treated herein and
does not suffer from the common degeneracies.

This image space is useful for general purposes such as exploring
a new survey by filtering, clustering, ordering and searching for morphologies
of interest, as demonstrated in Section \ref{applications}. All these analyses
and many others can be performed on the same image space, which may result 
in a great deal of time saving and parsimony for the astronomer. We approximate
the space to 12 dimensions because it is sufficient for our purposes, however in practice there is no need for the low rank approximation to be the same for all applications. For example, applications which depend on Euclidean distance benefit from lower dimensionality because the measure breaks down in high dimensions. Meanwhile, applications implicitly reliant on reconstruction fidelity which do not have issues with high dimensionality (such as missing data prediction) may benefit from approximating less by retaining more dimensions.

\subsection{Probabilistic interpretation}\label{ppca}

There are many applications which require some way of taking the
likelihood of galaxy points into account. This section describes a
formal equivalence between fitting PCA (an optimal low-rank
approximation) and fitting a Gaussian latent factor model under
certain conditions. This equivalence will enable us to interpret PCA
probabilistically and therefore to assign likelihood to galaxy
points.

Given data vectors from some $d$ dimensional space, a linear latent
factor model (LFM) aims to discover the basis for an optimal
projection to a $q<d$ dimensional space, usually under Gaussian
assumptions. In its simplest form the model can be written as
follows:

\begin{equation} \label{factormodel}
  t=Wx+\mu+\epsilon
\end{equation}

Here $t\in \mathbb{R}^d$ is the data vector, $x \in \mathbb{R}^q$ is
the latent vector, $W$ is a $d\times q$ projection matrix  relating
$t$ to $x$ where $q<d$, $\mu$ is an offset and $\epsilon$ is a
residual error. Usually it is given that $x\sim N(0,I)$, and that
$\epsilon \sim N(0,\Psi)$ where the form of $\Psi$ is to be
defined. This given, the properties of the normal distribution imply
that $t\sim N(\mu,WW^T+\Psi)$. \cite{whittle1952principal} showed that
in the case that $\Psi=\sigma^2 I$ (i.e. the covariance matrix is
diagonal and isotropic), and $\sigma^2$ is known, the maximum
likelihood estimation of the matrix $W$ is equivalent to the linear
least squares solution. Resultantly, $W$ spans the same subspace
as PCA and hence is also an optimal solution to the low rank
approximation problem formalised in equation \ref{mineq}.  However,
the formulation in \cite{whittle1952principal} is highly limiting
since it is unlikely that in real data the covariance structure is
entirely known or that the model and sample covariance are exactly the
same. However, \cite{tipping1999probabilistic} show that maximum
likelihood (ML) estimates for $W$ and $\sigma$ do exist without
requiring the covariance to be known and, that the scaled principal
eigenvectors make up the columns of $W$ when the estimators are at
their global maximum. They term this result probabilistic PCA
(PPCA). The result (explained more formally below) is that given the PCA
low rank approximation performed in the previous sections, we can
directly write down an equivalent factor model which induces a
multivariate Gaussian distribution over the image space and hence we
are able to assign likelihoods in that space to every galaxy
point. Other than the direct applications of this formulation covered
in Section \ref{applications}, the factor model allows us to compare
image spaces by quantifying the implications of their structural
differences on likelihood assignment to galaxy points, which we later
leverage to create representative samples.

We will now signpost the crucial points in the derivation from
\cite{tipping1999probabilistic}, but the interested reader is
encouraged to consult the paper directly. Given equation
\ref{factormodel} and the assumption of diagonal and isotropic error
$\epsilon\sim N(0,\sigma^2I)$, it follows that $t$ conditional on $x$
is given by $t\mid x \sim N(Wx+\mu,\sigma^2 I)$. Since $x\sim N(0,I)$
it is easy to marginalise over $x$ to obtain $t\sim N(\mu,
WW^T+\sigma^2 I)$. The corresponding log likelihood function is given
by:

\begin{equation}\label{loglik}
L=-\frac{N}{2}\Big[d\ln(2\pi)+\ln|C|+tr(C^{-1}S)\Big]
\end{equation}

where $S$ is the sample covariance matrix. The ML estimator for $\mu$
is the sample mean. Meanwhile, globally optimal estimates
for $\sigma,W$ can be obtained using iterative maximisation algorithms
such as those given in \cite{rubin1982algorithms}. Most importantly, what the authors show in \cite{tipping1999probabilistic} is that these
parameters can be obtained analytically using the artefacts from
PCA. The PCA equivalent of the $\sigma^2$ ML estimate is given by:

\begin{equation}
  \sigma^2_{ML}=\frac{1}{d-q}\sum_{j=q+1}^d \lambda_j
\end{equation}

where $\lambda_j$ are the excluded eigenvalues, hence it can be
roughly interpreted as the variance lost averaged over the number of
dimensions lost. The PCA equivalent of the $W$ ML estimate is given
by: 

\begin{equation}
  W_{ML}=U_q(\Delta_q-\sigma^2I)^{\frac{1}{2}}R
\end{equation}

were $\Delta_q$ is a $q\times q$ diagornal matrix with the retained
eigenvalues $\lambda_1,...,\lambda_q$ on its diagonal. $R$ is an
arbitrary rotation matrix and can be dropped for our purposes by
setting $R=I$. Thus, it is the case then that having calculated PCA,
we can use $\hat{\mu},\sigma_{ML},W_{ML}$ to immediately write down a
multivariate Guassian which induces a probability distribution over
the image space.

\begin{figure}
  \centering
  \includegraphics[scale=.65]{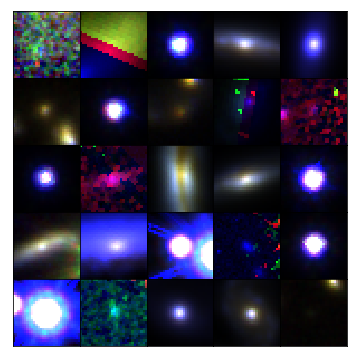}
  \includegraphics[scale=.65]{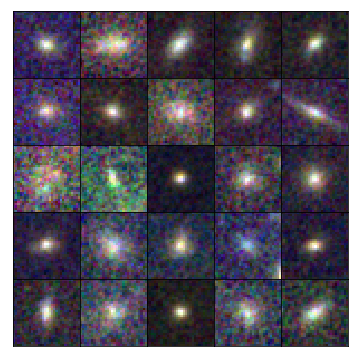}
  \caption{Each 1.8$''$ thumbnail in both images is an RGB composite
    of the F160W, F125W, and F814W bands. The top image shows $25$ of
    the least likely galaxies. The image shows anomalous detections
    and artefacts along with real, but intrinsically rare types of
    objects. The bottom image shows $25$ of the most likely
    galaxies. The image shows mainly poorly resolved galaxies with
    flux magnitude concentrated in a pronounced bulge, likely
    well represented by the first $2$ eigengalaxies. }
  \label{fig:galaxyorder}
\end{figure}

It is unlikely that the image space is Gaussian, not least
because it is unlikely to be symmetrical. However, this method for
assigning likelihood is still very useful because it provides a
consistent way to measure a galaxies distance from the mean average
galaxy with respect to the covariance of the data in the low-rank
approximated space. We can get some sense of the implications of this
likelihood assignment by considering the least and most likely
galaxies together. Figure \ref{fig:galaxyorder} illustrates the $25$
least likely (top) and the $25$ most likely (bottom) galaxies. The top
image shows anomalous detections and artefacts along with real, but
intrinsically rare types of objects. The bottom image shows mainly
poorly resolved galaxies with flux magnitude concentrated in a
pronounced bulge, likely corresponding to the first two
eigengalaxies.

\section{Applications}\label{applications}

In this section we explore some applications of the probabilistic
eigengalaxy framework described above.

\subsection{Outlier detection}

A key utility of outlier detection is to make discoverable rare
phenomena buried in enormous datasets. This may include searches for
exotic galaxies and rare objects but also the identification of
anomalous detections and pipeline errors. The two-part challenge is
first to define an `outlier' in a way useful to astronomy, and the
second is to scale the detection algorithm to the size of the data.

\citet{dutta2007distributed} implement a distributed version of PCA
using random projection and sampling to approximate eigenvectors for
the purpose of outlier detection and apply it to the 2MASS
\citep{skrutskie2006two} and SDSS \citep{fukugita1996sloan}
datasets. In defining an outlier they search for galaxies which are
over-represented by the last eigenvector. Other large scale PCA
methods include incremental PCA \citep{ross2008incremental}, which
approximates PCA by processing data in batches commensurate with the
available random access memory. \cite{baron2017weirdest} utilise a
random forest and fit to discriminate between real and synthetic
data\footnote{These authors use the flux at each wavelength as a
  feature set, generating synthetic data by sampling from the marginal
  distribution of each feature.}. For every pair of objects, they
count the number of trees in which each pair is labelled `real' in the
same leaf. The output is a $N\times N$ similarity matrix which is then
searched for `outliers', defined as objects with a large average
distance from all other objects.

We focus on a simple and principled definition for an `outlier',
proceeding directly from our eigengalaxy framework. Given the
probabilistic interpretation, a natural definition for an outlier is
an object with a low likelihood assignment which implies that it is
far from the mean given the covariance structure. Note that this same
approach could be applied not only to imaging, but also to
spectroscopy, light curves, catalogues and other kinds of data. We
define a formal outlier description within the eigengalaxy framework:
given a generative factor model $N(Wx + \mu,\sigma^2I)$, an outlier
$x$ is an object such that $p(x\mid Wx + \mu,\sigma^2I)<T$ where $T$
is a threshold likelihood, and can be set according to the purpose at
hand.

To illustrate the concept with the GZ-CANDELS dataset, we use our
derived eigengalaxies to assign a log likelihood to every galaxy,
using the {\it score\_samples} method on the {\tt
  sklearn.decomposition.PCA} object which implements the
\cite{tipping1999probabilistic} factor model representation to
calculate likelihood. We sort the data by likelihood and present in
the 25 objects with the lowest likelihood assignment. The result is
identical to that given in the top image of
Figure~\ref{fig:galaxyorder} which shows not only anomalous detections
and artefacts but also systems that are known to be rare, such as dust
lanes which are signposts of recent minor mergers \citep[see
  e.g.][]{Kaviraj2012b}, ongoing mergers \citep[see e.g.][]{Darg2010}
and edge-on spirals which appear to be accreting a blue
companion. Outlier detection therefore offers an efficient way to
identify examples of rare objects in large surveys.

\subsection{Predicting missing data}

\begin{figure}
\centering
\includegraphics[scale=0.5]{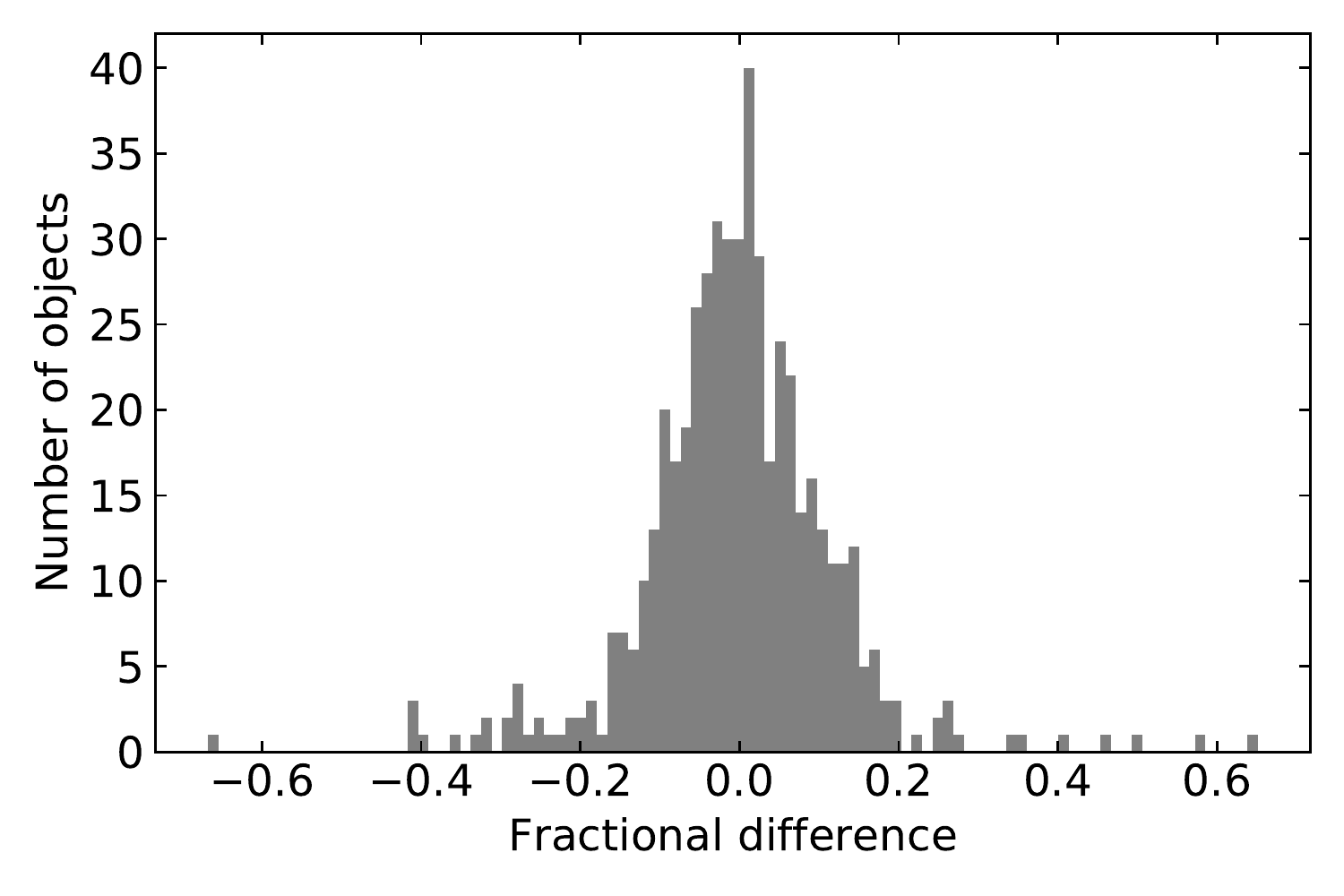}
\caption{The frequency of the ratio of the sum of deltas (difference
  between the original and predicted pixel flux) and the sum of the
  flux of the original images for galaxies with missing data. The body
  of the distribution extends to $\pm 20\%$ error but $\sim 72\%$ of
  objects are predicted within $\pm 10\%$ error, indicating that PPCA
  can predict missing data in a 2D sense with a high level of
  fidelity.}
\label{fig:recovery-rates}
\end{figure}

\begin{figure*}
\centering
\includegraphics[width=0.29\textwidth]{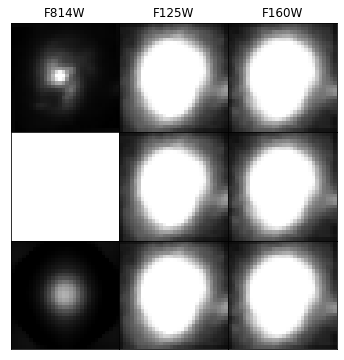}
\includegraphics[width=0.29\textwidth]{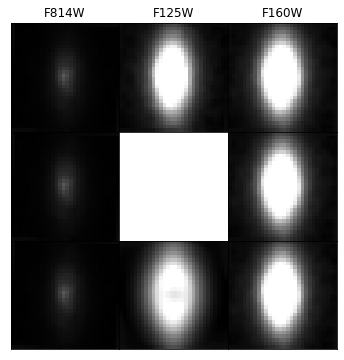}
\includegraphics[width=0.29\textwidth]{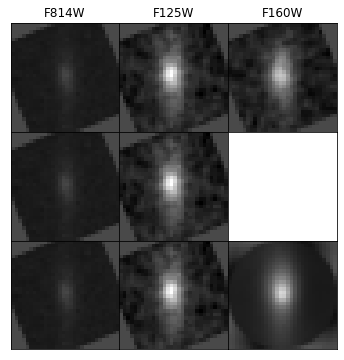}
\caption{An example of using PPCA for image imputation. In each
  figure, the first row shows the original galaxy across all three
  bands, the second row shows the same galaxy with a random band
  censored, the third row shows the censored galaxy predicted by
  PPCA. It is noteworthy that by using information contained in the
  eigengalaxies, PPCA is successful in estimating brightness correctly
  even when it bares no resemblance to that of the other bands, as
  illustrated in the left most image.}
\label{fig:impute-error}
\end{figure*}

In many situations one might be missing a particular band, for example
due to bad data, partial coverage with a certain bandpass,
obliteration by {\it Starlink} satellite trails, etc. In these cases
we can consider how well we can predict the missing data using
eigengalaxies. \cite{tipping1999probabilistic} define an expectation
maximisation (EM) algorithm for PPCA in the presence of missing data
which works by treating missing values as jointly distributed with the
latent variables and maximising the expectation of the joint
likelihood function. As a demonstration, we randomly omit a band with
equal probability from our dataset for 5\% of rows chosen at
random. We use the data as processed in Section 2.2 and an efficient
variational EM version of the \cite{tipping1999probabilistic}
algorithm\footnote{Available in Python package {\tt pyppca}.} set out
in \cite{porta2005active} to fit our PPCA model. We use the delta sum
over flux sum ratio as in Figure~\ref{fig:flux-diff-hist} for the
reconstruction error to describe the prediction
error. Figure~\ref{fig:recovery-rates} illustrates the distribution of
prediction error. The distribution is roughly symmetrical and is
centred on zero. The body of the distribution extends to $\pm 20\%$
error but $\sim 72\%$ of objects are accounted for within $\pm 10\%$
error showing that high fidelity prediction is possible for most
objects. Figure~\ref{fig:impute-error} provides some examples of
predicted data for different bands. It is noteworthy that by using
information contained in the eigengalaxies, PPCA is successful in
estimating total flux even when it bears no resemblance to that of the
other bands. The ability to predict missing images offers a route to
`filling in' missing data, such as predicting photometric data points
which are absent in the observations in order to reconstruct missing
parts of a galaxy's spectral energy distribution. 

\subsection{Searching for galaxies similar to an exemplar}

Given a survey with a large number of objects with diverse variety,
and an interest in galaxies of a specific kind, it is useful
to be able to present an exemplar galaxy and use it to quickly search
for all other galaxies with similar features. The utility and
suitability of a similarity search for any particular use case will
depend primarily on how the objects are being described, and how the
similarity between their descriptions is being calculated. For
example, \cite{protopapas2006finding} use cross-correlation as a proxy
for similarity between light curves for the purpose of detecting
outliers whereby light curves with the lowest average similarity are
defined as outliers. \cite{sart2010accelerating} utilise dynamic time
warps \citep{berndt1994using} to measure the similarity between light
curves. \cite{Hocking2018} compare various measures, including
Euclidean distance and Pearson's correlation coefficient, and use cosine distance to measure similarity from a description
generated by growing neural gas prior to hierarchical clustering. 

\begin{figure*}
\centering
\includegraphics[width=0.245\textwidth]{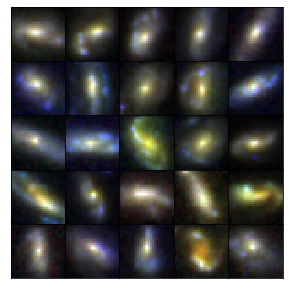}
\includegraphics[width=0.245\textwidth]{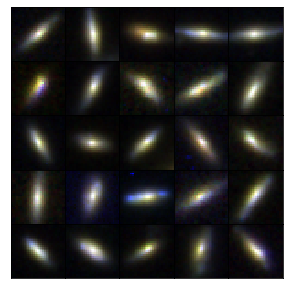}
\includegraphics[width=0.245\textwidth]{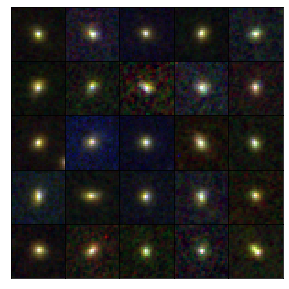}
\includegraphics[width=0.245\textwidth]{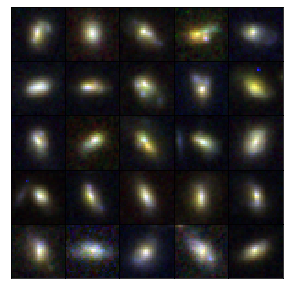}

\caption{Examples of similarity searches. In each 1.8$''$ thumbnail
  image (an RGB composite of the F160W, F125W, and F814W bands), the
  exemplar galaxy is given in the top left, followed by 24 of its
  nearest neighbours by Euclidean distance in the 12D eigengalaxy
  space. The image shows searches for spirals, edge-on spirals,
  ellipticals and recent mergers from left to right respectively. }
\label{fig:similar}
\end{figure*}

In Section \ref{lowrank}, we explained that the suitability of the image
space can be inspected by considering the similarity of the nearest
neighbours for any given galaxy, and we provided some examples in
Figure \ref{fig:similar-sample}. `Similarity search' works on precisely the
same principle. For any given reference galaxy, the rest of the galaxies
are ordered according to their Euclidean distance from it in
the rank reduced space. One may then consider each galaxy in
order of similarity. The method is principled because it is
established in Section \ref{intro} that the dimensions of the
space are not arbitrary, that they are contrived to be relevant to
morphology and that the number of dimensions retained pertains directly to
how well similarity neighbourhoods are preserved in the image
space. The method is also straightforward to extend to the
case where multiple reference objects are desirable, by calculating for
each other object the median distance from the reference points and
then ordering as before. Figure~\ref{fig:similar} illustrates searches
for spirals, edge-on spirals, ellipticals and recent mergers. It shows
very good likeness in the nearest neighbourhoods as may be expected
given previous tests.

\subsection{Unsupervised clustering}

Unsupervised clustering is particularly useful for astronomy
datasets because it provides a method for investigating very large
collections of objects efficiently as long as the image space has been
credibly constructed. If the clustering method is effective at
grouping objects with similar features together, then one
can study a dataset by characterising its {\it morphological centres}
rather than examining each object separately. There is some precedence
for this in astronomy. For example, \cite{Martin2020} follow
\cite{Hocking2018} in using growing neural gas and hierarchical
clustering directly on pixel data to identify structurally distinct
clusters. \cite{almeida2010automatic} and \cite{almeida2013automated}
utilise {\it k}-means to classify spectra from SDSS into fewer base
types. \cite{valenzuela2018unsupervised} use {\it
  k}-medoids\footnote{In {\it k}-medoids, datapoints become cluster
  centres, unlike {\it k}-means where the cluster centre is not 
  necessarily correspondent to a data point.} to cluster and map
sequences of light curve segments to variational trees. 

The eigengalaxy framework enables a simple unsupervised clustering
treatment. We can create a distance matrix (an $N \times N$ 
matrix in which each cell indicates a distance between object $j$ at
row $j$ and object $i$ at row $i$) by calculating the Euclidean
distance between every galaxy. The distance matrix provides an input
for a broad range of unsupervised clustering algorithms. This provides
us with an opportunity to define and discover groups of galaxies based
on the similarity of their multi-band morphologies as encoded by the
eigengalaxies. Amongst the central problems in clustering is
how the similarity between objects is defined and how many clusters
there are. In our case, similarity is defined by the Euclidean distance
between objects, and it turns out that there is a robust way to
conceive the clustering problem such that the number of clusters is
automatically decided. Let $\{d_{i,j}\}$ be an $N \times N$ distance
matrix in which each cell indicates the Euclidean distance between
object $j$ at row $j$ and object $i$ at row $i$. The objective then is
to choose a set of exemplar objects such that the sum of the distances
between each object and its closest exemplar is minimised. Formally,
for some set of galaxy points $s\in S$:

\begin{equation}\label{clustereq}
  \min_{\{q_i\}_{i=1}^m}\Bigg( \sum_{s\in S} \min_i{d_{s,q_i}} + \sum
  d_{q_i,q_i}\Bigg),
\end{equation}

where $q_i,...,q_m\in S$ are $m$ examplars. The second summation acts
as a regulariser barring trivial solutions (such as picking every point
as its own exemplar), and regulating the number of
exemplars that are chosen. The result is the selection of a set of objects
which we may call `cluster exemplars' and clustering is achieved by
labelling objects according to their closest exemplar. The benefit of
this formulation is that it is exact and will result in both the
number of centers and their membership. The exact problem is
NP-hard (i.e. cannot necessarily be solved in reasonable time) 
\citep{komodakis2009clustering} but there are myriad 
relaxations and approximations that deliver near optimal solutions in
empirical tests. One such approximation is known as affinity
propagation \citep{dueck2009affinity} (AP) which minimises
a similar equation to equation \ref{clustereq} using message passing 
over factor graphs.

\begin{figure*}
\centering

\includegraphics[width=0.245\textwidth]{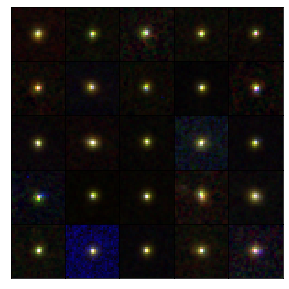}
\includegraphics[width=0.245\textwidth]{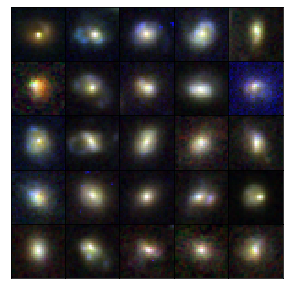}
\includegraphics[width=0.245\textwidth]{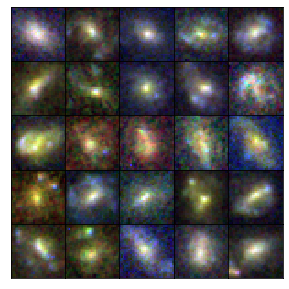}
\includegraphics[width=0.245\textwidth]{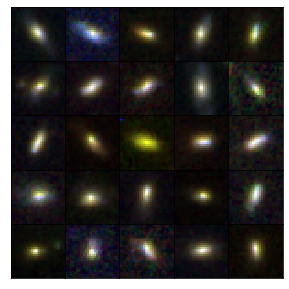}

\caption{Composite image samples of four morphological clusters from a
total of 536 created using affinity propagation clustering of a
distance matrix defined by pairwise Euclidean distances in 12D
eigengalaxy space. Each 1.8$''$ thumbnail image is an RGB composite of
the F160W, F125W, and F814W bands.}
\label{fig:ap}
\end{figure*}

To
demonstrate this, we produce a distance matrix for the GZ-CANDELS
dataset as outlined above. We use AP which produce 536 clusters from
the total sample of 10,243 objects, with clusters varying in size from
1 to 316 with a median size of 4. Affinity propagation has a tunable
`preference' parameter that allows the coarseness of clustering to be
adjusted; here we use the default value (the median of the distance
matrix). Figure~\ref{fig:ap} illustrates samples of galaxies from four
morphological clusters. It is noteworthy that affinity propagation is
an exemplar-based algorithm, so that each cluster is actually
characterised by an exemplar galaxy. Thus, in this instance the whole
dataset is summarised by 536 exemplars (5\% of the sample) which
could, in principle, be sorted by rarity (see Section\, 3.1) and then
viewed as one large image. This massive reduction in the scale of the
full dataset highlights the efficiency gains that can be made by using
this method for exploring the extremely large imaging surveys of the future.

\subsection{Combining methods}

The methods described above utilise the same framework and can be
used together to create additional capabilities simply based on the
eigengalaxy weights in the score matrix. There are many possible
combinations but we highlight three:

\begin{itemize}
    \item Missing value prediction $\to$ all methods. Predicting missing
      band values, dead pixels, etc. can be used to `complete' data so
      that it may be considered on par with the rest of the data. It may
      then be used with any of the other methods. 
    
    \item Outlier detection $\to$ similarity search / clustering /
      classification. We may discover an interesting outlier (e.g.\ a
      rare object such as a gravitationally lensed galaxy) of which we
      would like to find many more examples (similarity search). We
      may want to cluster the outliers to determine a self-similar
      morphological clusters. We may also want to label a training set
      of outliers, train a classifier and then select only the
      outliers of interest.
    
    \item Clustering $\to$ missing value prediction. To make the
      prediction of missing values more accurate, we may first
      generate eigengalaxies using the available bands and use it to
      create a clustering. This produces a set of self-similar classes
      to which missing values have more in common than the whole
      population. For each cluster we could then separately derive
      eigengalaxies and predict missing values for objects within that
      cluster.
\end{itemize}

\subsection{Big data}\label{bigdata}

The GZ-CANDELS dataset is relatively small and can be easily
processed. In this section we consider problems and adaptations which
may be necessary to use these methods with extremely large datasets,
such as LSST \citep{Robertson2017}, {\it Euclid} \citep{Refregier2010}
and the Square Kilometre Array \citep{Weltman2020} which may contain
billions of observed objects. We suggest that the eigengalaxy framework
offers a novel solution to processing such big data by making it
possible to find outliers, search for and cluster objects using only
the reduced score matrix form of a dataset. For example, if LSST was
to provide a corresponding score matrix, eigengalaxies and galaxy
likelihoods, all the methods described here would be available based
on a small fraction of the full dataset.

An important consideration in making the methods presented here
feasible for big data is the number of eigengalaxies required to
achieve a practically useful level of explained
variance. \citet{aggarwal2001surprising} provide a deeper discussion
on common issues for algorithms operating in high dimensional
space. In our case we would expect the following problems given too
many eigengalaxies:

\begin{enumerate}
    \item The score matrix form of the data may itself be too large to
      be useful. In our case, as the number of eigengalaxies
      approaches 10,243, the total size of the data converges onto the
      total size of the cut-out collection.

    \item Euclidean distance loses interpretability and the power to
      distinguish objects as the number of dimensions grows.
    
    \item Some techniques like fully calculating a distance matrix or
      calculating the similarity of every object with reference to an
      exemplar would not be computationally scalable.
\end{enumerate}

Taking LSST as an example, we would expect that the number of
eigengalaxies required to explain, say, 96\% variance to be more than
in our GZ-CANDELS experiments for two primary reasons: (i) it is a
deeper survey and it therefore observes more low mass galaxies, the
morphological mix of which is yet unknown \citep[e.g.][]{Martin2019},
and (ii) we would expect bright galaxies to have more detail such as
extended debris and tidal features \citep[e.g.][]{Duc2011} and
therefore exhibit more variance. However, the additional sources of
variance are limited, and we would expect a reasonable asymptotic
number of eigengalaxies to emerge for a useful level of explained
variance at relatively small sample sizes. 

The kind of variance that needs to be captured and the sufficient
ratio of it to retain ultimately depends on the intended
application. If the number of eigengalaxies is too high we could
explore various ways to reduce the variance not required for the
intended purpose in order to stay within an eigengalaxy quota at a
useful explained variance ratio. Potentially applicable techniques
include using smaller cut-outs, down-sampling pixels, using fewer
bands, combining bands, stretching the range of the flux densities to
make differences less subtle, and partitioning the data by some other
variable (e.g.\ brightness, location, etc.) and then conducting the
analysis per partition.

If the number of eigengalaxies was workable from a data size
perspective but a problem for Euclidean distance then we could use
other distance measures more robust in high dimensional
space. Finally, in every case we would need to replace exhaustive
similarity searches with more scalable methods such as the nearest
neighbour algorithm \citep{beis1997shape}, and affinity propagation
would need to be replaced with a less expensive clustering method such
as {\it k}-means (or {\it k}-medoids if Euclidean distance was
unacceptable).

\section{Conclusions and Summary} \label{conclusion}

We have demonstrated how galaxy cut-outs can be projected to a linear
similarity preserving image space. We used PCA for optimal low-rank
approximation and termed the resultant orthogonal basis vectors
`eigengalaxies'. We further showed how a formal equivalence of PCA to
latent factor models with isotropic variance enables a probabilistic
interpretation of the image space. This offers an empirical and
non-categorical approach to characterising galaxy morphology. Using a
sample of 10,243 galaxies from the Galaxy Zoo--CANDELS survey, chosen
to lie in the deep GOODS--S field, we use three-band {\it HST} ACS and
WFC3 imaging to derive just 12 eigengalaxies that are sufficient to
provide robust similarity preservation in the resultant image
space. We explored four applications of our framework:

\begin{enumerate}
    \item we have shown how the probabilistic interpretation can be
      used to assign likelihoods and identify outliers as objects with
      a low likelihood.
    \item we have shown how PPCA can be used to predict missing values
      in imaging, for example due to bad data or partial coverage in a
      given band. 
    \item we have shown how the projection of eigengalaxies on to a
      12D space naturally facilitates similarity searches, where
      galaxies can be sorted relative to their Euclidean distance from
      an exemplar, thus quickly returning samples of galaxies that are
      morphologically similar to the exemplar as defined by their
      eigengalaxy components. 
    \item we have shown how the Euclidean distances between galaxies
      in the 12D space can be compiled into a distance matrix that can
      provide the input for unsupervised clustering algorithms to
      discover groups of similar objects. We have demonstrated this
      using affinity propagation to show how morphologically similar
      groupings can be identified in large samples.
\end{enumerate}

We have described how the methods may be used together in
more advanced use cases and how working in eigengalaxy space may
present a novel solution to outlier detection, as well as search and
clustering problems in forthcoming massive imaging datasets such as
LSST. We argue that these results and illustrations underscore the
suitability of our PCA based probabilistic eigengalaxy framework for
the study of morphology, especially in the era of big data astronomy,
where representational efficiency and relevance will pay dividends.

\section*{Data availability}
Code for all results and figures can be found at
\url{https://emiruz.com/eigengalaxies}. 

\section*{Acknowledgements}
We warmly thank the referee for a thoughtful report that improved the
quality of the original manuscript. J.E.G. is supported by the Royal 
Society. S.K acknowledges a Senior Research Fellowship from Worcester College
Oxford. This work is based on observations taken by the CANDELS Multi-Cycle 
Treasury Program with the NASA/ESA HST, which is operated by the Association 
of Universities for Research in Astronomy, Inc., under NASA contract NAS5-26555.
This research made use of Astropy,\footnote{http://www.astropy.org} a 
community-developed core Python package for Astronomy \citep{astropy:2013, astropy:2018}.

\bibliographystyle{mnras}
\bibliography{eigen_v1}
\end{document}